%% file: paper.tex
\documentclass[acmsmall, nonacm]{acmart}

\usepackage{graphicx}
\usepackage{calc}
\usepackage{tikz}
\usepackage{multirow}
\usepackage[normalem]{ulem}
\useunder{\uline}{\ul}{}

\usetikzlibrary{calc, positioning, arrows.meta,
  decorations.pathreplacing, decorations.markings, shapes.multipart}
\newcommand{\outline}[1]{}
\newcommand{\Cfgblengines}{Configurable engines}
\newcommand{\cfgblengines}{configurable engines}
\newcommand{\cfgblengine}{configurable engine}
\newcommand\fmt{\ensuremath{\mathit{f\!mt}}}

\makeatletter
\DeclareFontEncoding{LS2}{}{\@noaccents}
\makeatother
\DeclareFontSubstitution{LS2}{stix}{m}{n}
\DeclareFontEncoding{LS1}{}{}
\DeclareFontSubstitution{LS1}{stix}{m}{n}

\DeclareSymbolFont{largesymbolsstix}{LS2}{stixex}{m}{n}
\DeclareSymbolFont{symbols2stix}{LS1}{stixfrak}{m}{n}

\DeclareMathDelimiter{\lbrbrak}{\mathopen}{largesymbolsstix}{"EE}{largesymbolsstix}{"14}
\DeclareMathDelimiter{\rbrbrak}{\mathclose}{largesymbolsstix}{"EF}{largesymbolsstix}{"15}
\DeclareMathDelimiter{\lBrace}{\mathopen}{largesymbolsstix}{"E8}{largesymbolsstix}{"0E}
\DeclareMathDelimiter{\rBrace}{\mathclose}{largesymbolsstix}{"E9}{largesymbolsstix}{"0F}
\DeclareMathSymbol{\langledot}{\mathopen}{symbols2stix}{"30}
\DeclareMathSymbol{\rangledot}{\mathclose}{symbols2stix}{"31}

\newtheorem{thm}{Theorem}

\begin{document}

\title{Comprehensive Verification of Packet Processing}

\author{Shengyi Wang}
\email{shengyiw@princeton.edu}
\orcid{0000-0002-2286-8703}
\author{Mengying Pan}
\email{mengying@princeton.edu}
\orcid{0000-0001-8970-9697}
\author{Andrew W.~Appel}
\email{appel@cs.princeton.edu}
\orcid{0000-0001-6009-0325}
\affiliation{%
  \institution{Princeton University}
  \city{Princeton}
  \state{New Jersey}
  \country{USA}
}
\authorsaddresses{December 27, 2024}

%% \authorrunning{S. Wang, M. Pan, and A. W. Appel}
\begin{abstract}
\textbf{Abstract.} To prove the functional correctness of a P4 program running in a
programmable network switch or smart NIC, prior works have focused
mainly on verifiers for the ``control block'' (match-action pipeline).
But to verify that a
switch handles packets according to a desired specification, proving
the control block is not enough.  We demonstrate a new \emph{comprehensive} 
framework for formally specifying and proving the additional components of the
switch that handle each packet: P4 parsers and deparsers, as well as
non-P4 components such as multicast engines, packet
generators, and resubmission paths.  These are generally triggered by
having the P4 program set header or metadata fields, which
prompt other switch components---fixed-function or configurable---to
execute the corresponding actions.  Overall behavior is correct only
if the ``configurable'' components are, indeed, configured properly;
and we show how to prove that. We demonstrate our framework by
verifying the correctness of packet-stream behavior in two classic P4
applications.  Our framework is the first to allow the
correctness proof of a P4 program to be composed with the correctness
proof for these other switch components to verify that the
switch programming as a whole accomplishes a specified behavior.

\end{abstract}

\maketitle              % typeset the header of the contribution

\section{Introduction}

\outline{The Significance of P4 Architectures}

High-speed network switches (and smart NICs) have many specialized, pipelined, limited-computation-depth
components.  Some of these components can be programmed in P4,
a domain-specific, hardware-independent \emph{dataplane programming}
language for high-speed processing of network packets.
P4 is designed to facilitate line-rate, single-pass, pipelined
processing of packets through a specialized switch fabric 
(though P4 can also be compiled to general-purpose CPUs); 
but this single-pass design naturally comes with 
significant limitations: P4 does not support general loops,
pointer data structures, or even (in some implementations) the ability
to access the same data structure more than once during the processing
of a packet.  The P4 ``pipeline'' comprises
a \emph{control block} (a.k.a. \emph{match-action pipeline})
sandwiched between
a \emph{packet parser} and a \emph{packet deparser}; a switch
might have more than one P4 pipeline, such as an
\emph{ingress pipeline} and an \emph{egress pipeline}.

Because of the language constraints imposed by the single-pass model,
network hardware (switches, NICs, etc.) also has non-P4 components
to handle specialized tasks. These components are accessed either
\emph{inside} the control block 
%\emph{during} the match-action pipeline 
%(as "extern" objects)
or \emph{before, between, and after} P4 pipelines; the P4 program can communicate
intentions to those components by setting fields in packet headers 
% that exit the P4 pipeline.
or calling "extern" functions.
These external components---such
as packet replicators, packet recirculators, multicast units, packet generators,
CPUs and other control-plane units---may be \emph{fixed function}
or they may be \emph{configurable}.  Collectively
we refer to them as \emph{\cfgblengines{}}.

Figure 1 shows the components of a Tofino switch.  In later sections we will explain their
functions; here just note that there are P4-programmable components
(purple, ``Ingress'' and ``Egress'' rows) and components that are ``configurable'' but not programmable
(grey, ``Traffic Manager'' row).
\begin{figure}[htb]
  \sffamily
  \centering
  \scalebox{0.7}{\input{paths.tex}}
  \caption{Packet Paths of the Intel Tofino Native
    Architecture. Purple boxes (Ingress and Egress) are the two P4 pipelines.
    Grey boxes (second row) are \cfgblengines{}.  }
  \label{fig:tofino}
\end{figure}
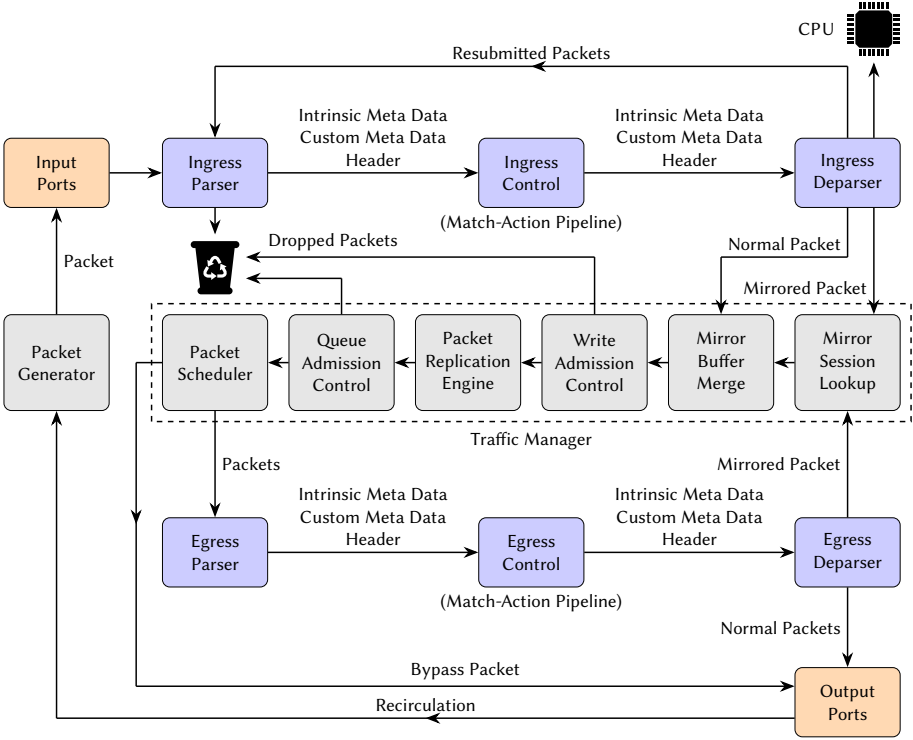

It's difficult to program and configure these switches, even using
an industry-standard language such as P4: the programming model
is (necessarily) different than conventional languages for CPU programming,
and testing the programs requires setting up a test network with packet
generators and high-speed switches. 
Therefore, formal program verification, to guarantee that a program+configuration
serves a specified purpose (or if there is a bug, to catch that bug) can be valuable.

In fact, there are some bugs that simply cannot be found by program testing.  In
one of our examples, the programmer had forgotten to initialize a field.  The compiler
is \emph{permitted} to initialize that field to zeros (in which case the bug
does not show up, which is what happened in this case); or is permitted to
initialize that field with arbitrary data, which would manifest as a bug,
but only with some future version of the compiler.
Program verification,
not against the compiler but against the \emph{specification} of the P4 language
and its interaction with all the non-P4 boxes, did indeed catch the bug---using 
the tools we describe in this paper.
Moreover, verifying the P4 program by itself
was not enough to find this bug; the bug
was detected when we tried to prove the end-to-end behavior
of our switch program+configuration.  What could (and did) detect
this is a formal proof quantifying over
all possible inputs and all legal compiler behaviors

\paragraph{Our goal} is to prove correctness of whole-switch programs/configurations
w.r.t. a formal specification of the desired behavior and w.r.t. a formal
model of the switch.  The difficulties are,
\begin{itemize}
  \item Previous verification tools focused almost exclusively on the
    P4 match-action control blocks, not other components,
    not even the P4-programmed parsers and deparsers  (see \S{}\ref{sec:relatework}).
  \item There were no high-level
formal models of the non-P4 components of the switch.
\end{itemize}

Without the ability to specify and verify all of these modules,
verification systems are incomplete: they cannot prove claims about
the overall behavior of a packet stream.

In this work, we make use of a previous P4 verification tool
(Verifiable P4).
Our new results are in specifying and verifying parsers, deparsers,
and \cfgblengines{}.  To write those specifications,
we have decoded the manufacturer's documentation
into a formal semantics on which we build formal tools.  Using those,
we demonstrate holistic verification of two example packet-processing
applications, a stateful firewall and a packet sampler, both
of which rely on \cfgblengines{} to realize their intended packet
behavior.
%(in addition to their core programming in P4)

In Related Work (\S\ref{sec:relatework}) we discuss
P4Testgen \cite{p4testgen}, which simulates an entire
Tofino switch---not only its P4 components but
also the configurable components we address in this paper.
It is a simulator, not a verifier, and
it is not based on a formal semantics---other than
its implementors' understanding of how the switch behaves,
as embodied in a C++ program that encodes queries for concolic execution.

Like our work, and unlike P4Testgen, 
HOL4P4 \cite{hol4p4_2024} formalizes a generic architecture semantics,
including both P4 and some \cfgblengines{} of a ``Very Simple Switch'' architecture
(which has a much simpler traffic manager than Tofino).
They prove correctness of a reference interpreter w.r.t. this
semantics, but do not show how to prove that P4 programs
(along with their configurable components) satisfy a
high-level specification.

% \outline{The Necessity of Verifying P4 Architectures}
% 
% In the same vein, to ensure a comprehensive verification of the packet
% processing procedure, it becomes essential to model the additional
% behavior, such as deparsers and, and integrate them with other verified P4 control blocks. This integration
% enables an end-to-end analysis of flow-level properties concerning
% packets. Additionally, it provides a valuable opportunity to validate
% the functional specifications of each block. The necessity for
% matching specifications---where the postcondition of one block
% indicates the postcondition of the subsequent block---facilitates the
% construction of a coherent and robust processing sequence. However,
% some existing frameworks, such as Verifiable P4 \cite{verifiablep4},
% which focus solely on the verification of P4 control blocks, lack this
% capability.

\outline{The Capabilities and Limitations of Verifiable P4}

\paragraph{The basis.}
Verifiable P4 \cite{verifiablep4} is a foundational interactive verification framework
designed for specifying and reasoning about data-plane packet
processing by P4 programs.  The framework includes a verifier
that provides program logic and automation support to prove,
for example, that a P4 program correctly implements the
functional model of an algorithm.
Unlike several P4 verifiers, it can handle P4 programs with
persistent state (i.e., a packet can influence
the processing of later packets), which is needed in some
applications.
Verifiable P4 is embedded in a general-purpose proof assistant
(Coq) in which one can then prove, for example, that the functional model
correctly satisfies a high-level specification on a packet stream.
The verifier is also
proven sound with respect to the operational semantics of P4.
%% Unfortunately, the program logic falls short of covering packet
%% parsing and deparsing, as well as additional behaviors. Addressing
%% packet deparsing is relatively straightforward, whereas packet parsing
%% and integrating additional behaviors require a more substantial
%% effort.

\outline{Our Framework for Verification of Whole Packet Processing}

\paragraph{Parsers/deparsers.}
Our new framework allows verification of the P4-programmable
packet-parsing (and packet-deparsing) functions, which turn
network packets (unstructured bit-sequences) into structured
data (and vice versa).  We refined the Verifiable P4
program logic to enable the verification of most common
packet parsers through an ad hoc approach.
Moreover, inspired by \textsc{Narcissus} \cite{narcissus}, we introduce a concise
notation to specify packet formats, which leads to a more
readable specification and more automated proofs.

%\paragraph{Non-P4 components of the network switch.}
\paragraph{\Cfgblengines{}.}
Our new framework contains formal models for several configurable components of the Tofino switch. For
components of interest with concrete behavior descriptions, such as
the packet generator and replication engine, we modeled them as
functions. For other components, we provided axioms to characterize the
minimal behaviors they must adhere to.

\paragraph{End-to-end verification.}
We show how to connect P4 verifications of parsers, control-blocks, and deparsers to lemmas about correct configuration of configurable engines,
to prove end-to-end properties of packet processing applications.

To demonstrate the strength of our framework, we verify two classic examples.
While this paper focuses on Intel's Tofino architecture,
the general methodology applies to other P4 architectures as well.

\subsection*{Contributions:}
\begin{enumerate}
%comprehensive
\item We introduce the concept of \emph{end-to-end data-plane verification
  for programmable switches,} grounded in the semantic
  specifications of each programmable, configurable, or fixed-function
  component.  This approach enables
  the validation of packet properties 
  from their entry into a network device to their exit.
\item We enhance the program logic of Verifiable P4 to support verification of packet
  parsers and deparsers. Additionally, we introduce a concise notation
  for specifying packet formats, leading to clear and readable
  specifications for packet parsers and deparsers. Our
  notation has a formal semantics, enabling us to prove the soundness of
  our verification system through a machine-checked
  proof.
\item We present formal models for several
  \cfgblengines{} within the Tofino switch. The packet
  replication engine is fully modeled, while we model useful subsets of the
  functionality for the packet generator and other components.
\item We demonstrate with two case studies: end-to-end verification of
  a packet sampler that integrates a P4 program proof with proofs of
  its correct use of \cfgblengines{} and parser/deparser; and improved
  verification of a stateful firewall that leverages \cfgblengine{}
  and parser specs to prove a stronger theorem than was previously
  possible.

\end{enumerate}

\section{Motivation}

\outline{The Rationale of Our Choice of Tofino and 2 Examples}

Intel
Tofino\texttrademark{} was
the first high-speed network switch to make
P4 programmability accessible
to general users.  It exemplifies the programming and programming-language
design choices for line-rate packet processing in a standards-compliant
domain-specific language.
In addition to its widespread use and accessibility, our choice of
Tofino as the modeling target is further justified by the many
\cfgblengines{} it offers. Figure~\ref{fig:tofino} illustrates various
\cfgblengines{}, such as packet generator, packet scheduler, packet
replication engine, and others in a Tofino switch.
A typical Tofino has four copies of this entire
pipeline, which do not share state with each other.

%\subsection{Packet Processing and Flow Control in Tofino}
\subsection{The Path of a Packet in the Data Plane}

Let us explore the journey of a packet in Figure~\ref{fig:tofino}
starting from the upper left corner. 
The packet is received at the Input Ports block---normally from outside the switch
but perhaps from inside the switch via ``recirculation'' or ``packet generation''.
Then the packet progresses
through subsequent blocks as shown by arrows. The packet first
enters the ingress pipeline, where the Ingress Parser parses the
packet headers, extracting metadata and segregating the payload.
The metadata indicates, for example, whether the packet came
from an external port or from recirculation/generation.
The Ingress Control block
handles data packet processing decisions, determining actions like
unicast, multicast, or dropping the packet by marking certain fields
in the metadata.\footnote{A packet arrives at the switch
with \emph{headers} and \emph{payload}.  Various switch components
add or modify headers and \emph{metadata}.  Finally,
outgoing packets depart with their modified headers and unmodified payload.}
Finally, the Ingress Deparser performs the opposite
function of the parser, reassembling the packet by prepending the
processed headers to the payload. If the Ingress
Deparser invokes a resubmit operation, the packet
goes back to the Ingress Parser instead of to the traffic manager.

If a packet is not resubmitted, it advances to the traffic manager,
passing through a sequence of five \cfgblengines{} as shown in
the second row of
Figure~\ref{fig:tofino}.
The packet may be duplicated (``mirrored''---see \S\ref{subsec:mirror}).
Then the Write Admission Control determines whether the packet
(and the duplicate, if any)
should be stored in a buffer pool or dropped.
The packet may then undergo replication
in the Packet Replication Engine, where the packet
is enqueued for multicast to multiple destinations
(see \S\ref{subsec:pre}).
Then the Queue Admission Control decides whether to enqueue or discard
each packet based on its priority metadata
and queue occupancy. Finally, the
Packet Scheduler apples various scheduling algorithms to dequeue 
%Packet Scheduler selects the appropriate queue from which to dequeue a
packets and
populates the egress intrinsic metadata.

If the \verb|bypass_egress| field in the metadata is not set by the
ingress pipeline, the packet proceeds to the egress pipeline. In the
egress pipeline, the Egress Parser, Egress Control, and Egress
Deparser undertake operations analogous to those in the ingress
pipeline. Additionally, the
Egress Deparser has the capability to send an egress-to-egress
mirrored packet to the traffic manager.

The configuration of the Traffic manager may
have \emph{port masks} which,
if matched by the packet headers and metadata,
will trigger packet recirculation in the
Packet Generator instead of forwarding to an output port.
The Packet Generator can also create
packets based on a periodic timer (or on other triggers;
see Section~\ref{subsec:pktgen}).

This
examination of potential packet flows illustrates that \cfgblengines{}
are crucial to the complete functionality of packet processing.

\begin{table}[htbp]
  \centering
  \scalebox{0.72}{
    \begin{tabular}{|c|c|c|c|c|c|l|l|}
    \hline
    \textbf{\cfgblengine}                                                                 & \textbf{PSA}         & \textbf{PNA}         & \textbf{Tofino}      & \textbf{v1model}     & \textbf{DPDK}        & \textbf{Applications}                                                          & \textbf{Examples}                                                                                                          \\ 
 & \cite{psa-arch} & \cite{pna-arch} & \cite{p416} & \cite{v1model} & \cite{DPDK16} & & \\     \hline
    \textbf{\begin{tabular}[c]{@{}c@{}}Replication\\(Multicast)\end{tabular}}             & Yes                  & Yes                  & Yes                  & Yes                  & No                   & Packet duplication                                                             & \begin{tabular}[c]{@{}l@{}}workload for performance test \cite{hypertester},\\ distributed network control \cite{lucid}\end{tabular} \\ \hline
    \multirow{2}{*}{\textbf{Recirculation}}                                                & \multirow{2}{*}{Yes} & \multirow{2}{*}{Yes} & \multirow{2}{*}{Yes} & \multirow{2}{*}{Yes} & \multirow{2}{*}{Yes} & \begin{tabular}[c]{@{}l@{}}Extra stateless \\ computation stages\end{tabular}  & \begin{tabular}[c]{@{}l@{}}secure hash computation \cite{smartcookie}, \\ loopback in service chain logic \cite{dejavu}\end{tabular} \\ \cline{7-8} 
                                                                                          &                      &                      &                      &                      &                      & \begin{tabular}[c]{@{}l@{}}Extra stateful \\ memory accesses\end{tabular}      & \begin{tabular}[c]{@{}l@{}}updating bandwidth limit policy \cite{ahab}, \\ second chance for hash collision \cite{dart}\end{tabular} \\ \hline
    \multirow{2}{*}{\textbf{Mirroring}}                                                    & \multirow{2}{*}{Yes} & \multirow{2}{*}{Yes} & \multirow{2}{*}{Yes} & \multirow{2}{*}{Yes} & \multirow{2}{*}{Yes} & \begin{tabular}[c]{@{}l@{}}Notification in \\ distributed systems\end{tabular} & \begin{tabular}[c]{@{}l@{}}logging in data centers \cite{bedrock},\\ distributed telemetry system \cite{spidermon}\end{tabular}      \\ \cline{7-8} 
                                                                                          &                      &                      &                      &                      &                      & \begin{tabular}[c]{@{}l@{}}Extra time within \\ the data plane\end{tabular}    & \begin{tabular}[c]{@{}l@{}}multitask measurement packet \cite{flymon}\\ in-network copy for messages\cite{redplane}\end{tabular}     \\ \hline
    \textbf{Resubmission}                                                                  & Yes                  & Yes                  & Yes                  & Yes                  & No                   & \begin{tabular}[c]{@{}l@{}}Extra time within \\ the ingress\end{tabular}       & \begin{tabular}[c]{@{}l@{}}loopback in service chain logic \cite{dejavu},\\ update TCP flow state \cite{blink}\end{tabular}          \\ \hline
    \multirow{2}{*}{\textbf{\begin{tabular}[c]{@{}c@{}}Packet\\Generator\end{tabular}}} & \multirow{2}{*}{No}  & \multirow{2}{*}{No}  & \multirow{2}{*}{Yes} & \multirow{2}{*}{No}  & \multirow{2}{*}{No}  & \begin{tabular}[c]{@{}l@{}}Out-of-band \\ traffic control\end{tabular}         & \begin{tabular}[c]{@{}l@{}}delayed network control \cite{lucid}, \\ time synchronization messages \cite{ripple}\end{tabular}         \\ \cline{7-8} 
                                                                                          &                      &                      &                      &                      &                      & \begin{tabular}[c]{@{}l@{}}Fast workload \\ generation\end{tabular}            & \begin{tabular}[c]{@{}l@{}}programmable load generator \cite{slog}, \\ workload for evaluation \cite{siconinp4}\end{tabular}       \\ \hline
    \end{tabular}
  }
  \caption{Configurable Engines in Different Architectures}
  \label{tbl:engines}
\end{table}

%% justify our choice of a particular architectures and two examples.
\subsection{Two Motivating Examples}

Numerous P4 network applications of interest rely on a variety of
\cfgblengines{} for their proper functionality.
Table~\ref{tbl:engines} demonstrates that architectures tend to share certain
common \cfgblengines{}. This observation suggests the feasibility and benefits of
modeling these \cfgblengines{} independently and integrating them with
P4-programmable components in a modular way. Such an approach
will streamline the verification of packet processing
applications across platforms.

To demonstrate our framework, we selected two P4 applications to prove the correctness of their intended behavior. The
first is a \textbf{packet sampler} that copies selected header fields from every 1024th packet
\label{example:sampler}
and sends them to a monitoring channel.
%It is implemented as follows.
%The ingress parser
%accepts and parses any TCP or UDP packet.  
To keep track of the number of packets
encountered, the ingress control block
uses a stateful counter, $c$. When processing a packet, if $(c + 1) \mod 1024 = 0$,
a new ``sample header'' with the sampled fields is added to the packet,
and a multicast group ID is assigned in the metadata. 
%The ingress
%deparser then reassembles the packet with the possibly modified header
%and metadata. 
The packet then proceeds to the traffic manager,
where (if a multicast group ID is set) it is duplicated.
%The egress parser processes the packet(s) from the
%traffic manager, whether or not they contain a ``sample
%header''.  
Now both the original and duplicated packets contain the sample header.
Therefore, in the egress pipeline, the control block will delete either the original headers or the 'sample' header, depending on whether the packet is a duplicate. This ensures that the original packet is restored and sent to its intended destination, while the sampled data is properly monitored.
%The egress
%deparser functions the same way as the ingress deparser.

The second example is an enhancement version of the \textbf{stateful firewall} example,
\label{example:firewall}
a P4 program verified for correctness by Wang \emph{et al.} \cite{verifiablep4}.
The firewall sits between a local area network and a hostile internet.
Each outgoing packet is assumed to be a ``request'' and its destination address
is put into a table (Bloom filter).  Each incoming packet is assumed to be hostile
and is dropped, unless it is  response to a request, determined by looking
up its source address in that same table.
To delete stale requests, the table needs to be cleaned regularly and incrementally each 10 milliseconds.
Wang \emph{et al.} proved the program
correct only subject to the assumption of a "dense flow", with no gaps between packets more than 10 milliseconds. 
With \cfgblengines, specifically the packet generator, we can periodically inject an extra packet into the flow every 10 ms
(thus provably producing a dense flow).
Therefore, we can prove correct a stateful firewall, comprising a P4 program (for the Ingress Pipeline) and
packet-generator configuration,  without having to assume
a dense flow arriving at the input ports.

\paragraph{In the remainder of the paper,}
we present our new work on specification and verification of P4 packet parsers
(\S\ref{sec:parser}),
modeling the \cfgblengines{} (\S\ref{sec:configurable}),
and how to put all these pieces together
into a comprehensive specification and proof of how
the entire switch (data plane) processes each packet, or
a stream of packets (\S\ref{sec:framework}).
Then we discuss related work and conclude.
% Molly: do we want to mention deparser at least somewhere given that we promised that in the intro

\section{Verifying Packet Parsers}
\label{sec:parser}

\outline{The Significance of Parser Verification}

Ensuring the correctness of a parser is crucial, as parsing serves as
the first line of defense against untrusted data for any software that
interacts with the outside world \cite{secappformal}.  
Failure to filter out adversarial
or malformed packets leads to increased vulnerability to network
attacks, potential system breaches, and compromised data
integrity.
One way to build secure packet parsers 
is from declarative grammar specifications \cite{bangert2014nail}.
Even better is to build \emph{provably} correct-by-construction
packet parsers, \emph{\`a la} \textsc{Narcissus} \cite{narcissus}.
For P4-specific reasons we will explain,
we have not quite achieved correct-by-construction,
but we have a declarative packet-grammar language
(inspired by \textsc{Narcissus})
for the specification and verification of parsers written in
P4.

\subsection{The Difficulty of Verifying P4 Packet Parsers}

Verifying a P4 packet parser presents its own unique set of
challenges, stemming from the variability of parser state transitions,
the complexity of possible architecture-specific behavior, and the
need to handle arbitrarily complex expressions and statements in
a reliable manner. Numerous ingenious techniques exist for verifying
parsers in general, and specifically for network protocols (see
\S\ref{sec:relatework} for further discussion). However, these
techniques are either correct-by-construction generators for
purely syntactic formats (that don't require semantic processing
mixed in with parsing)
or are limited to a restricted selection of
programming language primitives. None of them adequately meets our
requirements, because of the way P4 parsers mix syntactic
and semantic processing within the same parsing 
actions.\footnote{\label{langsec}It is generally considered inadvisable
to mix syntactic processing with semantic processing,
in the fields of compilers \cite[\S4.2]{appel98:mcij}
or packet parsing \cite{secappformal}.  In fact, the ``\textsc{LangSec}'' approach
to packet-processing security \cite{secappformal}
says, use syntactic-only correct-by-construction
methods to parse the packet, and \emph{do not start} the
semantic processing until the entire packet has (syntactically) parsed.
By enforcing this discipline, one avoids security attacks in which
ill-formed packets can ``trick'' the semantic analyzer to do bad things
the parser realizes that the latter part of the packet is syntactically incorrect.
But a P4 program running in a limited-depth pipelined network switch has
constraints that may make it impossible to follow this discipline;
neither in the P4 language nor in the switch's registers are there
the resources to store an abstract-syntax data structure.
So we say, if there are resource constraints that force the
programmer to do the dangerous mixing of syntactic with semantic
processing, the programmer had better prove \emph{formally} that
semantic processing of bad packets cannot cause
any nontrivial state changes.
We present in this paper the means for building
and checking that proof.}

Essentially a P4 parser is a P4 program describing an extended finite state automaton
with one \verb|start| state and two final states, \verb|accept| and
\verb|reject|.
Unlike in a traditional
finite state automaton, P4 programmers can write almost any
ordinary P4 statements within a state, as long as the last one is the
transition statement.

This
suggests that the verifier should treat each state as a normal
function in P4 with a special end: the transition statement is treated
as a function call.
But there are a few special considerations:
\begin{enumerate}
\item The P4 parsing program needs built-ins to access and advance through
  the packet headers, and our verifier must support those data extraction methods.
  The P4 core library includes a built-in \verb|extern| type named
  \verb|packet_in|, representing incoming network packets. A parser may
  invoke the \verb|extract| and \verb|advance| methods of a
  \verb|packet_in| type argument. We provide and prove the function
  specifications of these methods, directly from their existing
  operational semantics defined in Verifiable P4.
\item 
Finite state diagrams can contain loops (cycles in the FSA),
which also means the possibility of recursive function calls. 
The P4 language for Control programming (Ingress Control, Egress Control)
does not permit recursive functions \cite[Chapter 10]{p416},
since it is intended to target switch-fabric pipelines of finite depth,
and neither does the Verifiable P4 program logic.
But (in hardware such as Tofino) P4 parsers are implemented in
a diffent part of the chip that does permit loops via
iteration, up to 256 iterations.  Bounded recursions
can be verified in our framework but a nontrivial bound such as 256
would be unwieldy.  In future work it would be best to extend
the Verifiable P4 semantics with better support for bounded recursion.
\item Upon reaching a final \verb|accept| or
  \verb|reject| state, in the handoff of the parsed packet
  to the P4 match-action pipeline,
  the device's behavior is architecture-specific, not
  necessarily expressible in core P4; for verification one
  needs a semantic plug-in.\footnote{Some 
  P4 architectures provide default behavior,
  in the \texttt{reject} state, of dropping the packet.  In the
  Tofino architecture \cite{tofino-arch},
  if the ingress-control P4 program (which follows the parser)
  accesses the \texttt{parser\_err} field in the metadata, then
  the ingress control decides what to do with the packet;
  otherwise then the Tofino drops the packet
  before it reaches the ingress-control.  This can be expressed
  as a very simple program transformation.}

\end{enumerate}

The Verifiable P4 system already provides for modular verification
of modular P4 programs.  So we represent these ``plug-ins'' as
ordinary Verifiable P4 function specifications of \verb|extern|
methods, with preconditions and postconditions describing
how they access and modify the external state.

In this work,
we do not support testing \verb|parser_err|
in the (Ingress/Egress) Control;
therefore we can make the simplifying
assumption that error packets are dropped automatically.
This is sufficient for most P4 applications.
Also, we do not support (in the parser) calling the
\verb|reject| method---our semantic plug-in for \verb|reject| has
a precondition of \verb|False|.
Therefore, the way to reject packets is simply
not to provide a match-action transition for them;
this is the usual way that P4 parsers
reject packets.
Following the Tofino architecture,
our semantic plug-in for \verb|accept|
simply forwards packets to the
P4 ingress control pipeline without modification.
Consequently, our specification for \verb|accept|
has a precondition of \verb|True|.

\subsection{Approach and Limitation}

Existing tools for correct-by-construction parsing can
clearly and expressively handle the \emph{grammatical}
issues of parsing, but cannot fully handle the \emph{semantic actions}.
And the usual clean solutions---separate the parsing
phase from the semantic-action phase by abstract-syntax-tree constructors
(in separate passes), or in a single pass use higher-order action combinators---do
not work well in P4, with the lack of support
for AST data structures or higher-order functions.
So our approach is: We provide tools to prove that the P4 program---which
is state-machine-based (with P4 functions representing states)
and with semantic actions written in P4---correctly parses the grammar.
And we use Verifiable P4's existing
framework to verify correctness of the semantic actions.

Transition statements within a state come in two forms:
direct transitions to a state and \verb|select| expressions that
evaluate to a state.
Since the Verifiable P4 program logic does not yet support \verb|select|,
we define its semantics by transforming the
\verb|select| expression into an if-elsif chain, and reuse
the existing semantic and program-logic support for the \verb|if|
expression.

\subsection{Improvement of the Specification}

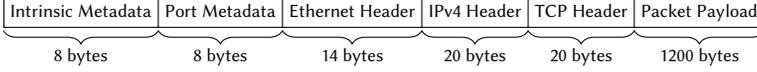
\begin{figure}[htbp]
  \centering
  \begin{tikzpicture}
    [format/.style={rectangle split, rectangle split parts=#1, draw,
        rectangle split horizontal, rectangle split part align=base},
      measure/.style={decorate,decoration={brace, mirror, raise=1pt,
          amplitude=5pt}}, cmt/.style={midway, below=5pt}]
    \sffamily
    \scriptsize
    \node [format=6] (packet) {Intrinsic Metadata\nodepart{two}Port
      Metadata\nodepart{three}Ethernet Header\nodepart{four}IPv4
      Header\nodepart{five}TCP Header\nodepart{six}Packet Payload};
    \draw[measure] (packet.south west)--(packet.one split south) node[cmt] {8 bytes};
    \draw[measure] (packet.one split south)--(packet.two split south) node[cmt] {8 bytes};
    \draw[measure] (packet.two split south)--(packet.three split south) node[cmt] {14 bytes};
    \draw[measure] (packet.three split south)--(packet.four split south) node[cmt] {20 bytes};
    \draw[measure] (packet.four split south)--(packet.five split south) node[cmt] {20 bytes};
    \draw[measure] (packet.five split south)--(packet.south east) node[cmt] {1200 bytes};
  \end{tikzpicture}
  \caption{A Typical Packet Byte Stream Format}\label{fig:packet}
\end{figure}

Even though correct-by-construction is not
a viable option (because those techniques cannot handle
the arbitrary semantic-processing code that may \emph{sometimes}
be needed in P4 parsing), 
we maintain our commitment to conciseness and elegance
through the design of a succinct notation for specifying packet
formats for \emph{verification}. A typical network packet prior to parsing in the Tofino
architecture might appear as shown in Fig.~\ref{fig:packet}.
Each segment of metadata and headers in the network packet has
distinct subformats that require parsing. Moreover, there are dependencies
between these segments. For instance, as illustrated in
Fig.~\ref{fig:packet}, the value of the protocol field in the IPv4
header dictates whether the subsequent header is TCP, UDP, or another
type. The primary mechanism for parsing a header of type \texttt{T}
(e.g., Ethernet, IPv4, TCP, UDP) in P4 is the \verb|extract| method,
with an essential specification as follows:
\begin{equation}\label{eqn:extract}
  \textsc{Extract}(\texttt{T},p) = (v, \textsc{Success}, p')
\end{equation}
where \textsc{Extract} represents the semantic function that extracts
a structured value $v$ of type \texttt{T} from a plain bit stream $p$,
and $p'$ denotes the remaining bit stream. For packet
formats like in Fig.~\ref{fig:packet}, the parser
specification might comprise a sequence of equations like
(\ref{eqn:extract}), each with a instantiated type \texttt{T}. This
approach can be lengthy and reveals numerous intermediate result
variables. The presence of dependencies further complicates the
matter, requiring branching statements in the specification.

Inspired by Delaware \emph{et al.}~\cite{narcissus}, we developed a format
supporting accurate matching, underspecification, concatenation, and
branching; and we defined a predicate $p\vDash \fmt$
to assert that a packet $p$ satisfies a
format $\fmt$.

For example, the following format specification permits either IPv4 packets (as shown in Figure~\ref{fig:packet}) or IPv6, and handles TCP or UDP:
\begin{gather*}
  \begin{split}
    p\vDash & \lbrbrak \mathsf{meta} \rbrbrak; \langle 64 \rangle;
    \lbrbrak \mathsf{ethernet} \rbrbrak; \lbrbrak \mathsf{ipv4}
    \rbrbrak; \\ & \lBrace \text{is\_tcp}\,\mathsf{ipv4} \,?\,
    \lbrbrak \mathsf{rslt} \rbrbrak \mid \lBrace
    \text{is\_udp}\,\mathsf{ipv4} \,?\, \lbrbrak \mathsf{rslt}
    \rbrbrak \mid \varepsilon \rBrace \rBrace; \langledot p'
    \rangledot
  \end{split}\\
  \begin{aligned}
    \mathsf{meta}
    \VDash_{\mathsf{T}}&\,\mathsf{meta\_h}\\ \mathsf{ethernet}
    \VDash_{\mathsf{T}}&\,\mathsf{ethernet\_h}\\ \mathsf{ipv4}
    \VDash_{\mathsf{T}}&\,\mathsf{ipv4\_h}
  \end{aligned}\qquad\quad
  \mathsf{rslt} \VDash_{\mathsf{T}}
  \begin{cases}
    \mathsf{tcp\_h} & \text{is\_tcp}\,\mathsf{ipv4}\\
    \mathsf{udp\_h} & \text{is\_udp}\,\mathsf{ipv4}\\
    \mathbb{0} & \text{otherwise}
  \end{cases}
\end{gather*}
where $v \VDash_{\mathsf{T}} T$ means the value $v$ has P4 type
$T$, and $\mathbb{0}$ is the empty type. Equation~\ref{eqn:format} shows
the internal syntax for packet formats:
\begin{equation}\label{eqn:format}
  \begin{aligned}
    \fmt \mathrel{\mathop:}=&\,\varepsilon & \text{empty}\\
    \mid &\, \lbrbrak v \rbrbrak & \text{exact encoding match}\\
    \mid &\, \langledot p \rangledot & \text{exact plain match}\\
    \mid &\, \fmt_1 ; \fmt_2 & \text{concatenation}\\
    \mid &\, \lBrace b \,?\,\fmt_1  \mid \fmt_2 \rBrace & \text{branching}\\
  \end{aligned}
\end{equation}
This new format does not reveal the intermediate parsing results and
properly encodes the dependencies. The matching rule is encoded in the
definition of predicate $\vDash$.
\emph{Exact encoding match} $p \models \lbrbrak v \rbrbrak$
means that the bit stream $p$ is exactly the encoding of the
structured value $v$, and \emph{exact plain match} $p \models \langledot p' \rangledot$ means
$p = p'$, where $p'$ is an unparsed bitstream.

%% \subsection{Automated and semiautomated verification of parsers...}

%% write something to go here \ldots

\paragraph{Proving parsers.}  To conveniently prove in Coq that parsing programs written in P4 satisfy specifications written with the $\vDash$ judgment, we built Coq definitions, lemmas, and tactics to mostly automate the task.

\section{Modeling the Configurable Engines}
\label{sec:configurable}
Another piece missing from most P4 verifiers is modeling the
\cfgblengines{} present on a typical switch or smart NIC.  Typically,
all P4-programmable components, including parsers, controls, and
deparsers, adhere to a fundamental design principle of ``one packet
plus metadata in, one packet plus metadata out.''  In order to drop,
replicate, or resubmit a packet, the P4 program must assign values in
headers or metadata as signals to these \cfgblengines{} that actually
perform the action.  To verify the switch's overall behavior, one
needs operational models of those \cfgblengines{}.

As presented in Fig.~\ref{fig:tofino}, each gray box represents
a configurable engine endowed with specific
behaviors. For clarity and focus, our modeling effort concentrates on
the two essential components highlighted in
our examples: the packet replication engine and the packet
generator. We adopted a minimal modeling approach for other
components, including mirror session lookup, write admission control,
queue admission control, and the packet scheduler,
modeling only the default configuration of each engine,
which allows packets to pass through these components unchanged.
The intricacies of some components, such as mirror
session lookup
or the handling of packet priorities in queue admission
control, may be explored in future research. On the other hand,
the packet scheduler remains underspecified in our study due to the
absence of publicly available documentation on the packet scheduling
algorithm employed in the Tofino architecture.

Our model is grounded
in the publicly available description of Tofino \cite{tofino-arch}.
It represents an
under-approximation, primarily because we have opted to exclude
certain details  (such as the \emph{packet descriptor})
that are primarily aimed at optimizing space
utilization and enhancing performance. Given that these aspects are
not within the scope of our concern, we have deliberately omitted such
details from our model.

\subsection{Modeling the Packet Replication Engine}\label{subsec:pre}

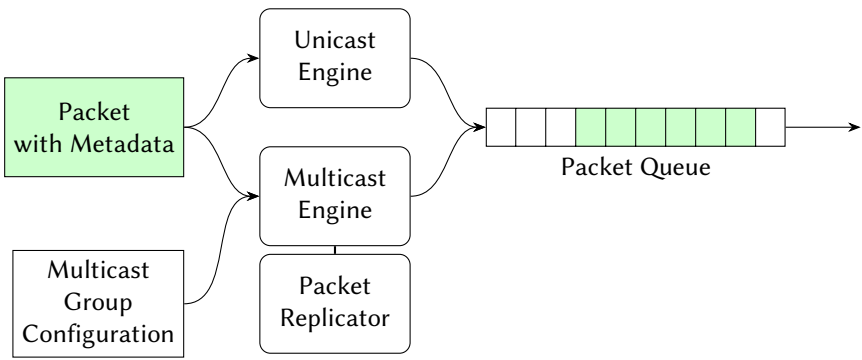
\begin{figure}[htb]
  \sffamily
  \centering
  \input{pre.tex}
  \caption{Packet Replication Engine}
  \label{fig:pre}
\end{figure}

The Packet Replication Engine (PRE) creates copies
of a packet, handling both unicast and multicast.
As indicated in Table~\ref{tbl:engines}, the PRE is a
ubiquitous component across various architectures, yet its specific
behavior is closely tied to the architecture in question. To address
this, we have developed a concrete model of the PRE tailored to the
Tofino architecture, aiming to strike a balance between universality
and architectural specificity. This approach is designed to simplify
the process of adapting our PRE model for use in different
architectural contexts.

\outline{What does a PRE do?}

\begin{figure}[htb]
  \sffamily
  \centering
  \input{groupconfig.tex}
  \caption{A Multicast Group Configuration}
  \label{fig:groupconfig}
\end{figure}
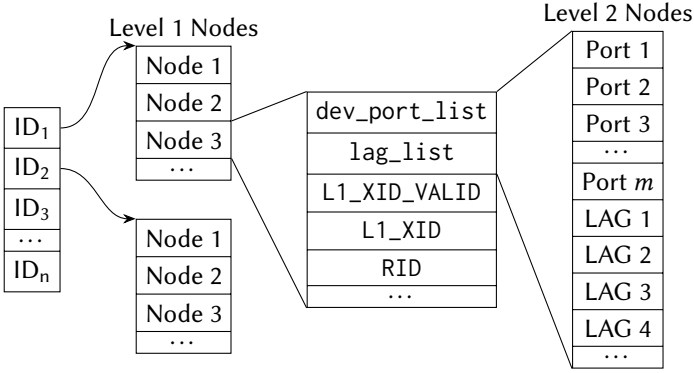

For a single packet, a Tofino PRE examines various fields within its
metadata to determine if replication is necessary.
The output of this engine is a sequence of packets to be enqueued
by the next stage of the Traffic Manager.

The method of
replication is 
dictated by a 2-level table lookup
(illustrated in Fig.~\ref{fig:groupconfig}), in a table configured
by the control plane.
The input to this table lookup is the \emph{metadata},
a structured bitstring calculated by the Ingress Control
and/or by Mirror Session Lookup.
There are four fields of interest: a port field
(\verb|ucast_egress_port|) that determines if a unicast
copy will be made; a 1-bit flag (\verb|copy_to_cpu|) indicating
whether to create a copy to send to the CPU; and two fields
(\verb|mcast_grp_a| and \verb|mcast_grp_b|) representing multicast
group IDs.  Looking up an ID in the table yields
to a specific configuration for
multicast replication.

We model the PRE as with a function for the multicast engine and another for unicast:
\begin{gather*}
  \textsc{MulticastEngine}(c:\mathrm{mc\_config})(p: \mathrm{ingress\_packet}) : \mathrm{list}(\mathrm{egress\_packet}) := \\
     \ldots  \mbox{\textit{details elided, but see Figure~\ref{fig:groupconfig}}}\\
~\\  
\textsc{UnicastEngine}(p: \mathrm{ingress\_packet}) : \mathrm{option}(\mathrm{egress\_packet}) :=  \\
     \ldots  \mbox{\textit{details elided}}\\
~\\ 
\textsc{ReplicationEngine}(c,p,\vec{p}')~ :=\\
\vec{p}' = \textsc{MulticastEngine}(c)(p) \cdot \textsc{UnicastEngine}(p)
\end{gather*}

Our definitions of \verb|multicast_engine| and \verb|unicast_engine|
are fully precise, derived from the semiformal specifications in the Tofino documentation.
The configuration $c$ comprises a multicast table and an L2 exclusion table.
Here, the types ingress\_packet and egress\_packet are really just the packet metadata,
since the packet headers and payload are neither read nor written by the Traffic Manager.

Based on this behavioral specification,
the user can take a configuration and---in
a larger context we will describe in the next section---prove
in Coq that a Tofino switch configured this way implements the desired end-to-end
packet behavior.

\subsection{Modeling the Packet Generator}\label{subsec:pktgen}

In Tofino, each pipeline (i.e., set of components shown in Figure~\ref{fig:tofino})  is equipped with a packet generator capable of
producing up to 8 independent streams of packets. 
These streams are configured and enabled
by the control plane. The control plane software can specify one of
four event types to trigger the packet generation process: one-time
timer, periodic timer, port down, and packet recirculation. For
simplicity, we model only the periodic timer type of packet generator
and a single stream of packet generation.

Regardless of the event type, once the specified event occurs, it will
initiate the creation of one or more batches of packets, with each
batch consisting of a configured number of packets. Therefore, a set
of parameters is required, including the number of batches, the time
interval between batch starts, the number of packets per batch, the
time interval between packets, etc. We bundle all these configuration
settings into a single configuration parameter and define a periodic
packet generation function accordingly:

\begin{equation*}
  \textsc{PktGen}(c: \mathrm{pktgen\_config})(t: \mathbb{Z})(s_\mathrm{g}: \mathrm{pktgen\_state}):=\dots
\end{equation*}

Here, the unit of time is the switch's clock tick, so time progresses
discretely. The time-related parameters in the configuration $c$ are
much larger than 1. We assume that \textsc{PktGen} is invoked at every
tick. When appropriate, it will generate a packet, but each time it
will also produce the new (possibly unchanged) state. Using
\textsc{PktGen}, we can define the axiom of a general packet
generator. Note that the output packet $p_\mathrm{g}$ may be invalid,
as the packet generator does not produce a packet at every time $t$.

\begin{gather*}
\mathbf{Axiom}~ \mathrm{PacketGenBehavior}: \hspace{3in}\\
 \textsc{PacketGenerator}(c, t, s_\mathrm{g},s'_\mathrm{g},p_\mathrm{recirc},p'_\mathrm{recirc},p_\mathrm{g)} ~\rightarrow \\
s_\mathrm{g}=s'_\mathrm{g}\wedge
\mathrm{valid}(p_\mathrm{recirc}) \wedge p_\mathrm{g}=p_\mathrm{recirc} \wedge \mathrm{invalid}(p'_\mathrm{recirc}) ~~\vee\\
\mathrm{invalid}(p_\mathrm{recirc}) \wedge \textsc{PktGen}(c, t, s_\mathrm{g}) = (p_\mathrm{g}, s'_\mathrm{g})
\end{gather*}

This axiom states that at every tick, either a recirculated or
generated packet is emitted, or no packet is emitted at all. Each
time, the packet generator state $s_\mathrm{g}$ transitions to a new
state, $s’_\mathrm{g}$.

\subsection{Modeling the Input Ports}\label{subsec:inputports}
The Input Ports are a complex multistream structure.  We can axiomatize
an underapproximation of its behavior, that it takes either a packet
$p_\mathrm{g}$ from the packet generator,
or it takes a packet (not necessarily the earliest available one) from
the input history $q_\mathrm{input}$,
and forwards this as $p_\mathrm{i}$ to the Ingress Pipeline.

\begin{gather*}
\mathbf{Axiom}~ \mathrm{SimpleInputPorts}: ~~
\forall p_\mathrm{g},q_\mathrm{input},q'_\mathrm{input},m_\mathrm{i},p_\mathrm{i},\\
\textsc{InputPorts}(p_\mathrm{g},q_\mathrm{input},q'_\mathrm{input},m_\mathrm{i},p_\mathrm{i})~\rightarrow\\
(\mathrm{invalid}(p_\mathrm{g})~\wedge~\exists q_\mathrm{l},q_\mathrm{r}, ~~q_\mathrm{input}=q_\mathrm{l}\cdot p_\mathrm{i} \cdot  q_\mathrm{r}  ~\wedge~q'_\mathrm{input}=q_\mathrm{l} \cdot  q_\mathrm{r})\\
\vee~~(q'_\mathrm{input}=q_\mathrm{input}\wedge p_\mathrm{i}=p_\mathrm{g})
\end{gather*}

\subsection{Modeling the Mirror Session Machinery}\label{subsec:mirror}

In some application one wants a single input packet to produce two different output packets,
with different headers, perhaps intended for different destinations.
The \emph{Mirror} machinery can accomplish that: the P4 program running in the
Ingress Pipeline marks the packet-metadata for mirroring, and then the traffic
manager implements the duplication.

In every clock tick, the Ingress Deparser produces one normal packet (possibly marked
as ``drop'') and one mirrored packet (marked as ``valid'' or ``invalid'', i.e., ``present'' or ``absent'')
with a Mirror Session ID.
Mirror Session Lookup uses the ID as an index in a table to provide metadata that
the remaining components of the traffic manager will use to route the packet.
In each tick, at most one packet flows from Mirror Session Lookup to the Mirror Buffer,
which is a queue of packets with metadata.
The Mirror Buffer flow is merged into the normal flow:
in each tick, at most one packet flows into
Write Admission Control; either a normal packet from Ingress Deparser or
(with lower priority) a packet from the Mirror Buffer.
This means that normal packets are never lost (in this merge) but mirror packets
can be lost.
In Figure~\ref{fig:tofino}
we indicate the combination of the buffer and the merging as Mirror Buffer Merge.

We specify the Mirror Session Lookup as a relation between input
(Mirrored Packet from the Ingress Deparser) and output.
At present, we have axioms to specify behavior of only the empty configuration of the mirror session tables,
and when the P4 program does not mark packets for mirroring.
The Mirror Buffer Merge simply propagates normal packets, and the Mirror Buffer queue stays empty.
In future work, we can axiomatize the nontrivial case.

\begin{gather*}
\mathbf{Axiom}~ \mathrm{EmptyMirrorTable}: \\
\forall c: \mathrm{config}, i: \mathrm{mirror\_session\_ID}, m: \mathrm{metadata}, \\
c=\mathrm{empty} ~\rightarrow~\textsc{MirrorSessionLookup}(c,i,m) ~ \rightarrow ~
\mathrm{invalid}(m)\\
~\\
\mathbf{Axiom} ~\mathrm{EmptyMirrorMerge}: \\
 \forall m_\mathrm{normal}, m_\mathrm{mirror}, q_\mathrm{old}, m_\mathrm{out},  q_\mathrm{new},\\
  q_\mathrm{old}=\mathrm{empty\_queue}~\wedge~\mathrm{invalid}(m_\mathrm{mirror}) ~\rightarrow \\
  \textsc{MirrorBufferMerge}(m_\mathrm{normal},m_\mathrm{mirror},q_\mathrm{old},m_\mathrm{out},q_\mathrm{new}) ~\rightarrow \\
  m_\mathrm{out}=m_\mathrm{normal}~\wedge~q_\mathrm{new}=\mathrm{empty\_queue}\\
\end{gather*}

\subsection{Modeling Queue Admission Control}
The Tofino traffic manager has a complex queue structure with multiple queues
for each output port, with a priority structure between queues.
The Queue Admission Control has a built-in mechanism to trigger flow-control packets sent
back to the sources of incoming packet streams, when queues get full.  For many applications,
it is sufficient to model a subset of this behavior.  The simplest possible axiomatic
characterization of this module is that some of the new packets (coming from the Replication Engine)
may be dropped, but packets already in the queue(s) are not lost.

The predicate 
$\textsc{QueueAdmissionControl}(\vec{m_\mathrm{repl}},p,q_\mathrm{egress},q\dag_\mathrm{egress})$
takes the list of replicated packets $\vec{m_\mathrm{repl}}$.  Actually,
all these packets share the same ``raw packet'' $p$ consisting of headers and payload
created by the Ingress Deparser; they differ only in their metadata, and
in fact $\vec{m_\mathrm{repl}}$ is just a list of differing metadatas created
by the packet replicator.

The weakest useful axiom characterizing Queue Admission Control is that 
some sublist of $\vec{m_\mathrm{repl}}$
is selected, the packet $p$ is attached to each one, and the resulting
metadata-equipped packets are appended to the tail of the egress queue.

\begin{gather*}
\mathbf{Axiom}~ \mathrm{MinimalQueueAdmissionControl}: \\
 \forall \vec{m_\mathrm{repl}},p,q_\mathrm{egress},q\dag_\mathrm{egress}, \\
  \textsc{QueueAdmissionControl}(\vec{m_\mathrm{repl}},p,q_\mathrm{egress},q\dag_\mathrm{egress})~\rightarrow \\
  \exists \vec{m'}, m' \subset \vec{m_\mathrm{repl}} ~\wedge~
    q\dag_\mathrm{egress} = q_\mathrm{egress} \cdot \mathrm{map}(\lambda m.(m,p))m'
\end{gather*}    

For some applications we want a stronger axiom to prove the application properties
of interest.  For example, we would like to prove that the packet sampler never
drops any packets, provided that the destination ethernet at the output port
is willing to accept them.
We can express that in a stronger axiom for Queue Admission Control.
We define an \emph{always-ready} ethernet as one that can accept a packet whenever
it arrives at the ethernet's output port,
and we say that packets destined for always-ready ports are not dropped:

\begin{gather*}
\mathbf{Axiom}~ \mathrm{AlwaysReadyQueueAdmissionControl}: \\
 \forall \vec{m_\mathrm{repl}},p,q_\mathrm{egress},q\dag_\mathrm{egress}, \\
  \textsc{QueueAdmissionControl}(\vec{m_\mathrm{repl}},p,q_\mathrm{egress},q\dag_\mathrm{egress})~\rightarrow \\
  \exists \vec{m'}, \vec{m'} \subset \vec{m_\mathrm{repl}} ~\wedge~\\
    (\forall m, m\in\vec{m_\mathrm{repl}} \wedge \textsc{AlwaysReady}(\mathrm{port}(m)) \rightarrow m\in \vec{m'}) ~\wedge \\
    q\dag_\mathrm{egress} = q_\mathrm{egress} \cdot \mathrm{map}(\lambda m.(m,p))m'
\end{gather*}    

This will allow us to prove that the packet sampler never drops a packet if the output port is always ready;
see Section~\ref{sampler-correct}.

\subsection{Modeling the Packet Scheduler}
When an output port is empty (because ethernet transmission of its prior contents has completed\footnote{more precisely, when it will complete within the number of clock ticks equal to the length
of the Egress Pipeline}), the Packet Scheduler can remove a packet from one of
the output queues destined for that port, and send it into the Egress Pipeline.
The simplest possible axiomatic
characterization of the Packet Scheduler is to model all the queues as a big bag of packets,
and that some packet is removed from the bag and forwarded.
In future work, we can model more configuration parameters for the scheduler
that control how multiple queues and priorities are handled in a way that can
guarantee, for example, that certain classes of packets are not reordered.

\begin{gather*}
\mathbf{Axiom}~ \mathrm{MinimalPacketScheduler}: \\
 \forall q,q',m,p,
  \textsc{PacketScheduler}(q,q',m,p) ~\rightarrow\\
  \exists q_\mathrm{l},q_\mathrm{r}, q=q_\mathrm{l}\cdot (m,p) \cdot  q_\mathrm{r} ~\wedge~ q'=q_\mathrm{l} \cdot q_\mathrm{r}
\end{gather*}    

\subsection{Modeling the Output Ports}
At each clock tick, the Output Ports may receive a packet
$p_\mathrm{e}$ from the Egress Pipeline or from the bypass mechanism;
in this work we have not modeled bypass packets.
The packet comes
with an indication $m_\mathrm{e}$ of whether it is
marked for Recirculation, in which case it goes to the Packet
Generator as indicated in Figure~\ref{fig:tofino}.
The variable $q_\mathrm{output}$ is not a queue, it is the history of packets transmitted from
the switch.

\begin{gather*}
\mathbf{Axiom}~ \mathrm{MinimalOutputPorts}: \\
 \forall q_\mathrm{output}, m_\mathrm{e}, p_\mathrm{e}, q'_\mathrm{output},p_\mathrm{recirc},\\
 \textsc{OutputPorts}(q_\mathrm{output},(m_\mathrm{e},p_\mathrm{e}),q'_\mathrm{output},p_\mathrm{recirc})~\rightarrow \\
 m_\mathrm{e}=0 \wedge q'_\mathrm{output}=q_\mathrm{output}\cdot p_\mathrm{e} \wedge \mathrm{invalid}(p_\mathrm{recirc})~\vee\\
 m_\mathrm{e}=1 \wedge q'_\mathrm{output}=q_\mathrm{output} \wedge p_\mathrm{recirc}=p_\mathrm{e}
\end{gather*}

\section{A Framework Connecting Them All}
\label{sec:framework}
\outline{The necessity of end-to-end verification}

Our goal is to formally characterize the whole behavior of the switch
when the P4 programs are installed in the programmable components
and configuration settings are installed in the configurable components.
Previous work \cite{verifiablep4} formally specified the
Ingress Control and Egress Control;
in Section \ref{sec:parser} we described formal specifications
of the Ingress and Egress Parsers and Deparsers;
in Section~\ref{sec:configurable} we described formal specifications
of the \cfgblengines{} (although some of those are underspecified,
i.e., specifying behavior of only some of the allowable configuration settings).
Each of those is specified in the form of an input-output state-transition relation
on packet flow.

In this section we will formally characterize the ``plumbing,'' described by the
lines and arrows in Figure~\ref{fig:tofino}.  That will allow an end-to-end
(input ports to output ports) behavioral characterization of a (programmed
and configured) Tofino switch.
We will do this directly in the Coq logic,
modeling each component as a stateful input-output relation.

\subsection{Connecting the P4 components together}
\paragraph{The P4 components} are the \emph{ingress pipeline} and the
\emph{egress pipeline}, and each connects
three subcomponents: the parser, the control block, and
the deparser.  No queues are needed between these
synchronously clocked, one-packet-in one-packet-out
subcomponents.
We can model each component as a relation
\begin{gather*}
R_c(d_\mathrm{in},s_\mathrm{old},d_\mathrm{out},s_\mathrm{new})\\
  c\in\{\textsc{InPrsr},\textsc{InCtrl},\textsc{InDprsr}, \textsc{EPrsr},\textsc{ECtrl},\textsc{EDprsr}\}
\end{gather*}
where $c$ is the name of the component,
$d_\mathrm{in},d_\mathrm{out}$ are the packet (with its metadata) entering and leaving,
the $s_\mathrm{old}, s_\mathrm{new}$ are the local state of each
component.  There is no global state accessible by multiple
components.

We specify a P4 pipeline such as the top row of Figure~\ref{fig:tofino} as,
\begin{gather*}
% d_\mathrm{in} : intrinsic_metadata \times port\_metadata \times raw\_packet
\textsc{IngressPipeline}(d_\mathrm{in}, s_\mathrm{old}, d_\mathrm{out},s_\mathrm{new}) := \\
  \exists s_\textsc{ip},s_\textsc{ic},s_\textsc{id}, ~ s_\mathrm{old}=(s_\textsc{ip},s_\textsc{ic},s_\textsc{id}) ~\wedge \\
  \exists s'_\textsc{ip},s'_\textsc{ic},s'_\textsc{id},~ s_\mathrm{new}=(s'_\textsc{ip},s'_\textsc{ic},s'_\textsc{id}) ~\wedge\\
  \exists t_1: \mathrm{timestamps}, p: \mathrm{raw\_packet}, d_\mathrm{in}=(t_1,p) ~\wedge \\
  \bigl(R_\textsc{InPrsr}(p,s_\textsc{ip},\textsc{None},s'_\textsc{ip}) ~ \wedge ~ s_\textsc{ic} = s'_\textsc{ic} ~
  \wedge ~ s_\textsc{id} = s'_\textsc{id} ~\wedge~ d_\mathrm{out} = \textsc{None} \\
  ~\vee \\
  \exists d_1: \mathrm{metadata}\times \mathrm{headers},  ~ l: \mathrm{payload}, \\
  R_\textsc{InPrsr}(p,s_\textsc{ip},\textsc{Some}(d_1,l),s'_\textsc{ip}) ~ \wedge \\
   \exists m_2:\mathrm{metadata\_for\_traffic\_manager}, d_2: \mathrm{metadata}\times\mathrm{headers}, \\
  R_\textsc{InCtrl}((t_1,d_1),s_\textsc{ic},(m_2,m_\mathrm{mirror},d_2),s'_\textsc{ic})~ \wedge \\
 \exists m_3: \mathrm{metadata}, ~h_3:\mathrm{headers}, \\
  R_\textsc{InDprsr}(d_2,s_\textsc{id},(m_3,h_3),s'_\textsc{id})~ \wedge~
  d_\mathrm{out} = \textsc{Some}((m_2,m_\mathrm{mirror},m_3), h_3\cdot l)\bigr) 
\end{gather*}

\noindent This plumbing is not quite as simple as shown in Figure~\ref{fig:tofino}.
For example, timestamps $t_1$ bypass the Parser,
the payload $l$ bypasses the Ingress Control and Deparser,
metadata $m_2$ bypasses the Deparser, but metadata within $d_2$
flows into the Deparser. And if $m_\mathrm{mirror}$ is a nonzero mirror session ID,
then (just after the Ingress pipeline),
a copy of the packet is to be sent to the Mirror Buffer.
We also consider another case: when a packet cannot be correctly
parsed, it is dropped. In this situation, it is not transferred to the
subsequent stages, and no output packet is generated in the pipeline.

% With the verifiable P4 parsers and deparsers, alongside the formal
% models of \cfgblengines{}, and leveraging existing verifiable P4
% control blocks, we can finally integrate them all to perform
% comprehensive end-to-end verification of packet processing. Just as
% end-to-end testing complements unit testing, end-to-end verification
% provides additional trustworthiness based on individually verified
% components. The latter are indispensable ingredients of the
% former. Without end-to-end testing, it is challenging to ensure that
% all units are interconnected correctly. In order for end-to-end
% verification to be feasible at all, the component verifications must
% be in a rich enough specification language to describe the overall
% desired behavior. Fortunately, Verifiable P4 has a very rich
% specification language that allows us to ensure that the
% specifications of individual components align and to prove the
% integrity of the entire system.
% 
% For \emph{foundational end-to-end verification,} that is, machine-checked
% proof from axioms and first principles, those axioms include
% the operational specification of each component (P4 match-action
% semantics, P4 parsing/deparsing semantics, semantics of 
% each \cfgblengine{})---which we have covered in the previous sections---and
% modeling the pipeline structure that connects all these components together.
% In a seamless connection, the postcondition of one component must
% satisfy the precondition of the next.

\paragraph{The egress pipeline}
is modeled in a similar way to the ingress pipeline, but it is simpler.
The pipeline receives metadata $m_\mathrm{e}$ and a packet $p_\mathrm{e}$,
and produces a modified packet $p'_\mathrm{e}$.
The output metadata $m'_\mathrm{e}$ is an indication of whether the
packet should be recirculated or output normally.
The pipeline has its own persistent registers whose state
$s_\mathrm{e}$ may be modified into $s'_\mathrm{e}$.

\begin{gather*}
  \textsc{EgressPipeline}((m_\mathrm{e},p_\mathrm{e}),s_\mathrm{e},(m'_\mathrm{e},p_\mathrm{e}'),s'_\mathrm{e}):=\\
  \exists s_\textsc{ep},s_\textsc{ec},s_\textsc{ed}, ~ s_\mathrm{e}=(s_\textsc{ep},s_\textsc{ec},s_\textsc{ed}) ~\wedge \\
  \exists s'_\textsc{ep},s'_\textsc{ec},s'_\textsc{ed},~ s'_\mathrm{e}=(s'_\textsc{ep},s'_\textsc{ec},s'_\textsc{ed}) ~\wedge\\
  \exists d_1: \mathrm{metadata}\times \mathrm{headers},  ~ l: \mathrm{payload}, \\
  R_\textsc{EPrsr}((m_\mathrm{e},p_\mathrm{e}),s_\textsc{ep},(d_1,l),s'_\textsc{ep}) ~ \wedge \\
   \exists d_2: \mathrm{metadata}\times\mathrm{headers}, ~
  R_\textsc{ECtrl}(d_1,s_\textsc{ec}, d_2, s'_\textsc{ec})~ \wedge \\
  \exists m_3: \mathrm{metadata}, ~h_3:\mathrm{headers}, \\
  R_\textsc{EDprsr}(d_2,s_\textsc{ed},(m_3,h_3),s'_\textsc{ed})~ \wedge~
  m'_\mathrm{e} = (m_2, m_3) \wedge p_\mathrm{e}' = h_3\cdot l
\end{gather*}

The real work here is done by the Egress Parser (modeled by relation 
$R_\textsc{EPrsr}$) the Egress Control (modeled by
$R_\textsc{ECtrl}$) and the Egress Deparser (modeled by $R_\textsc{EDprsr}$).

\subsection{Connecting all of the components}

As illustrated in Fig.~\ref{fig:tofino}, P4-programmable components
and \cfgblengines{} can be arranged in intricate configurations
designed for a particular architecture. Given that the configuration
is fixed for each specific architecture, it allows for the
creation of a comprehensive pipeline model that is uniquely adapted to
each architecture.  Here we will describe the ``plumbing'' of
the Tofino, and the same approach can be used to model other architectures.

In contrast to the P4 pipelines, which have no internal queues
and can be modeled as stateful relations on
individual (input and output) packets, 
we model the traffic manager as a relation on queue states
(and on the persistent states $s$ of the Ingress pipeline).

The state variables of the switch are,
\begin{description}
\item[$t$]  time measured in clock ticks, provided to the Ingress Control as a possibly useful piece of metadata,
\item[$s_\mathrm{g}$] state of the Packet Generator,
\item[$s_\mathrm{i}$] state of the persistent registers of the P4 Ingress Pipeline,
\item[$s_\mathrm{e}$] state of the persistent registers of the P4 Egress Pipeline,
\item[$q_\mathrm{input}$]  history of packets received by the switch,
\item[$p_\mathrm{recirc}$] recirculated packet (if any),
\item[$q_\mathrm{mirror}$] contents of mirror buffer,
\item[$q_\mathrm{egress}$] contents of output queues,
\item[$q_\mathrm{output}$] history of packets transmitted by the switch.
\end{description}

\begin{gather*}
  \textsc{IngressPacket} ((c_\mathrm{mirror},c_\mathrm{mc}, c_\mathrm{pktgen}): \mathrm{config}) \hspace{2in} \\
    (t,s_\mathrm{g},s_\mathrm{i},s_\mathrm{e}) (q_\mathrm{input}, p_\mathrm{recirc}, q_\mathrm{mirror}, q_\mathrm{egress}, q_\mathrm{output})\\
    (t',s'_\mathrm{g},s'_\mathrm{i},s'_\mathrm{e}) (q'_\mathrm{input}, p'_\mathrm{recirc}, q'_\mathrm{mirror}, q'_\mathrm{egress}, q'_\mathrm{output})
   := \\
  s_\mathrm{e}=s'_\mathrm{e}\wedge q_\mathrm{output}=q'_\mathrm{output} ~\wedge\\
  \exists p_\mathrm{g},m_\mathrm{i},p_\mathrm{i},m_\mathrm{normal},m_\mathrm{mirror},p'_\mathrm{i},m_\mathrm{merge},p_\mathrm{repl},\\
  \textsc{PacketGenerator}(c_\mathrm{pktgen}, t,s_\mathrm{g},s'_\mathrm{g},p_\mathrm{recirc},p'_\mathrm{recirc},p_\mathrm{g)}~\wedge~t'=t+1~\wedge\\
  \textsc{InputPorts}(p_\mathrm{g},q_\mathrm{input},q'_\mathrm{input},m_\mathrm{i},p_\mathrm{i}) ~\wedge\\
  \bigl(\textsc{IngressPipeline}((t,m_\mathrm{i},p_\mathrm{i}),s_\mathrm{i},\textsc{None},s'_\mathrm{i})~\wedge~
  q_\mathrm{mirror}=q'_\mathrm{mirror} ~\wedge ~ q_\mathrm{egress}=q'_\mathrm{egress}  \\
  \vee~
  \textsc{IngressPipeline}((t,m_\mathrm{i},p_\mathrm{i}),s_\mathrm{i},\textsc{Some}((m_\mathrm{normal},\mathit{id}_\mathrm{mirror}),p'_\mathrm{i}),s'_\mathrm{i})~\wedge\\
  \textsc{MirrorSessionLookup}(c_\mathrm{mirror},\mathit{id}_\mathrm{mirror},m_\mathrm{mirror})~\wedge\\
  \textsc{MirrorBufferMerge}(m_\mathrm{normal},m_\mathrm{mirror},q_\mathrm{mirror},m_\mathrm{merge},q'_\mathrm{mirror})~\wedge\\
  \textsc{ReplicationEngine}(c_\mathrm{mc},m_\mathrm{merge},\vec{m_\mathrm{repl}}) ~\wedge \\
  \textsc{QueueAdmissionControl}(\vec{m_\mathrm{repl}},p'_\mathrm{i},q_\mathrm{egress},q'_\mathrm{egress})\bigr). \\
\end{gather*}
\begin{gather*}
  \textsc{EgressPacket} (t,s_\mathrm{g},s_\mathrm{i},s_\mathrm{e}) (q_\mathrm{input}, p_\mathrm{recirc}, q_\mathrm{mirror}, q_\mathrm{egress}, q_\mathrm{output}) \hspace{1in}\\
    (t',s'_\mathrm{g},s'_\mathrm{i},s'_\mathrm{e}) (q'_\mathrm{input}, p'_\mathrm{recirc}, q'_\mathrm{mirror}, q'_\mathrm{egress}, q'_\mathrm{output})~:= \\
  \exists m_\mathrm{e},p_\mathrm{e},m'_\mathrm{e},p'_\mathrm{e},\\
  t=t'\wedge s_\mathrm{g}=s'_\mathrm{g} \wedge s_\mathrm{i}=s'_\mathrm{i}\wedge q_\mathrm{input}=q'_\mathrm{input} \wedge \mathrm{invalid}(p_\mathrm{recirc}) \\
  \textsc{PacketScheduler}(q_\mathrm{egress},q'_\mathrm{egress},m_\mathrm{e},p_\mathrm{e}) ~\wedge\\
  \textsc{EgressPipeline}((m_\mathrm{e},p_\mathrm{e}),s_\mathrm{e},(m'_\mathrm{e},p_\mathrm{e}'),s'_\mathrm{e})~\wedge\\
  \textsc{OutputPorts}(q_\mathrm{output},(m'_\mathrm{e},p'_\mathrm{e}),q'_\mathrm{output},p'_\mathrm{recirc}). \\
~\\
  \textsc{ProcessPacket}~ c~s~q~s'~q' := \textsc{IngressPacket}~c~s~q~s'~q' \vee
  \textsc{EgressPacket} ~s~q~s'~q'. \\
  \textsc{ProcessPackets}~ c~s~q~s'~q' := ~~s=s'\wedge q=q' ~\vee ~ \hspace{2in}\\
     \exists s'',q'', ~  \textsc{ProcessPackets}~ c~s~q~s''~q'' \wedge   \textsc{ProcessPacket}~ c~s''~q''~s'~q'.
\end{gather*}
The Ingress system (from Input Ports all the way to Queue Admission Control) fills up the
output queues (and drains the $p_\mathrm{recirc}$ register); it runs concurrently 
with the Egress system (from Packet Scheduler to
Output Ports) that drains these queues (and may fill $p_\mathrm{recirc}$).

\section{Proving correctness properties of applications}

The purpose of formally specifying the many components of the switch (or smart NIC) is to be
able to formally prove correctness of P4 applications running on the switch.

\subsection{The LangSec property}

Using the Verifiable P4 system \cite{verifiablep4} within Coq,
the user can prove that a specific P4 program refines a particular
$R_\textsc{InCtrl}$ relation.  Using our new work, the user can prove that
P4 programs refine particular $R_\textsc{InPrsr}$
$R_\textsc{InDprsr}$ relations.   Using our definition of \textsc{IngressPipeline}
one can compose these to derive
an overall behavior, $(R_\textsc{InPrsr},R_\textsc{InCtrl},R_\textsc{InDprsr})$.

Then one can prove that the Ingress behavior has the desired properties.  For example, the \textsc{LangSec} properties (see footnote \ref{langsec}):
\emph{Syntactically ill-formed packets are dropped with no effect on the state}.

\textbf{Lemma:} Our packet sampler parser is not affected by the parser state $s_\textsc{ip}$.
\begin{gather*}
  \textsc{ParserStateOblivious}(R_\textsc{InPrsr}) := \\
  \forall s_\textsc{ip1}, s_\textsc{ip2}, s'_\textsc{ip1}, s'_\textsc{ip2}, p, d_1,l_1,d_2,l_2, \\
  R_\textsc{InPrsr}(p,s_1,(d_1,l_1),s'_1) ~\rightarrow ~ \\
  R_\textsc{InPrsr}(p,s_2,(d_2,l_2),s'_2) ~\rightarrow ~ \\
  d_1 = d_2 ~\wedge~ l_1 = l_2 \\
\end{gather*}
\textbf{Proof:} Initialize
$R_\textsc{InPrsr}$ with the concrete parser specification of the
sampler example. From the specification, we find that the value of
metadata $d_i$ and payload $l_i$ only depends on the input packet
$p$. So we can prove $d_1 = d_2$ and $l_1 = l_2$ since they are parsed
from the same packet $p$.

\textbf{Theorem:} Our packet sampler has the LangSec property: a syntactically ill-formed packet
cannot affect the state.  In particular, if the \textsc{InPrsr} returns \textsc{None},
then $s_\textsc{ip}$ is changed in an oblivious way, and the rest of the
state $s_\mathrm{g},s_\mathrm{in},s_\mathrm{id},s_\mathrm{e}$ is not changed at all.

\begin{gather*}
  \textsc{LangSec}(R_\textsc{InPrsr},R_\textsc{InCtrl},R_\textsc{InDprsr}) :=\\
    \textsc{ParserStateless}(R_\textsc{InPrsr})~ \wedge\\
    \textsc{ProcessPacket}~ c~(t,s_\mathrm{g},(s_\mathrm{ip},s_\mathrm{in},s_\mathrm{id}),s_\mathrm{e})~(p\cdot q_\mathrm{input}, p_\mathrm{recirc}, q_\mathrm{mirror}, q_\mathrm{egress}, q_\mathrm{output})\\
    \hspace{2in} ~(t',s'_\mathrm{g},(s'_\mathrm{ip},s'_\mathrm{in},s'_\mathrm{id}),s'_\mathrm{e})~q' ~\wedge\\
      R_\textsc{InPrsr}(p,s_\textsc{ip},\textsc{None},s'_\textsc{ip}) ~ \rightarrow  ~ \\
      q' = (q_\mathrm{input}, p_\mathrm{recirc}, q_\mathrm{mirror}, q_\mathrm{egress}, q_\mathrm{output}) ~ \wedge
      s_\mathrm{g}=s'_\mathrm{g}~\wedge~s_\mathrm{in}=s'_\mathrm{in}~\wedge~s_\mathrm{id}=s'_\mathrm{id}~\wedge~s_\mathrm{e}=s'_\mathrm{e}.
\end{gather*}

\noindent

These are easy proofs, and they should be.  They are provable directly from the
\emph{specifications} of the P4 programs (that is, relations
$R_\textsc{InPrsr},R_\textsc{InCtrl},R_\textsc{InDprsr}$),
without having to look at the P4 code again (after proving that the P4 program satisfies these specifications).

\subsection{Proving the sampler correct.}
\label{sampler-correct}
We can also demonstrate a whole-switch correctness property:
the packet sampler correctly produces a digest of the
headers of every 1024th packet.
We describe this specification by relating the
output stream $q_\mathrm{output}$ to the input stream $q_\mathrm{input}$.

\begin{gather*}
  \textsc{SamplerSpecification}(n, q_\mathrm{input}, q_\mathrm{output})=\\
  \Big(\mathrm{empty}(q_\mathrm{input})\wedge\mathrm{empty}(q_\mathrm{output})\Big) \vee \null\\
  \Big(\exists p_\mathrm{input}, p_\mathrm{output}, q_1, q_2,~~
  q_\mathrm{input}=q_1 \cdot p_\mathrm{input} ~\wedge~
  q_\mathrm{output} \subset q_2 \cdot p_\mathrm{output} ~ \wedge \null\\
  (n + |q_1| + 1)\!\! \mod 1024 \neq 0 \wedge
  \textsc{NormalPacketRelation}(p_{\mathrm{input}}, p_{\mathrm{output}}) \wedge\null \\
  \textsc{SamplerSpecification}(n, q_1, q_2)\Big) \vee \null\\
  \Big(\exists p_{\mathrm{input}}, p_1, p_2, q_1, q_2,~~
  q_\mathrm{input}= q_1 \cdot p_\mathrm{input}~ \wedge ~
  q_\mathrm{output} \subset q_2 \cdot p_1 \cdot p_2~ \wedge \null \\
  (n + |q_1| + 1)\!\! \mod 1024 = 0 ~\wedge ~
  \textsc{NormalPacketRelation}(p_{\mathrm{input}}, p_1) \wedge \null \\
  \textsc{SpecialPacketRelation}(n, p_{\mathrm{input}}, p_2) \wedge 
  \textsc{SamplerSpecification}(n, q_1, q_2)\Big)
\end{gather*}
The sampler program synthesizes a special packet for every 1024th
packet. Each of the three clauses in the large disjunction expression
above corresponds to one possible case for a switch running the
sampler: the empty case, the case when a normal packet arrives, and when the
counter determines that 1024 packets have passed since the last
special packet. The \textsc{NormalPacketRelation} and
\textsc{SpecialPacketRelation} are predicates describing the
relationship between input and output packets.

Then we proved the following main theorem.
\begin{thm}[Sampler Correct]\label{thm:sampler}
\begin{gather*}
  \mathrm{counter}(s_\mathrm{i})=n \rightarrow \\
 \textsc{ProcessPackets}(c_\mathrm{sampler})
    (t,s_\mathrm{g},s_\mathrm{i},s_\mathrm{e}) (q_\mathrm{input},\mathrm{nil},\mathrm{nil},\mathrm{nil},\mathrm{nil})\\
    (t',s'_\mathrm{g},s'_\mathrm{i},s'_\mathrm{e}) (\mathrm{nil}, p'_\mathrm{recirc}, q'_\mathrm{mirror}, q'_\mathrm{egress}, q_\mathrm{output})
~~
 \rightarrow\\
 \textsc{SamplerSpecification}(n, q_\mathrm{input}, q'_\mathrm{output}).
\end{gather*}
\end{thm}
\noindent That is, when the switch is configured and programmed to run our sampler, then
if the counter value is $n$ in the initial
Ingress Control state $s_\mathrm{i}$,
then the relation between input stream $q_\mathrm{input}$ and output
stream $q_\mathrm{output}$ will indeed be as specified by \textsc{SamplerSpecification}.

%% \textbf{The rest of this section needs substantial adjustments}

\paragraph{Proof.}
The proof can be divided into two stages. Stage 1 involves proving two
relations \textsc{NormalPacketRelation} and
\textsc{SpecialPacketRelation} for a single pair of input and
corresponding output packets under different conditions. This is a
low-level proof, as it requires proving properties directly from the
P4 code based on its semantics. Stage 2 involves proving
\textsc{SamplerSpecification} for the entire input and output
queues. This stage is a relatively high-level proof, as it abstracts
the low-level properties into two pairwise relations and conditions.

In Stage 1, we consider the two predicates \textsc{IngressPacket} and
\textsc{EgressPacket} separately. Essentially, only the
\textsc{IngressPipeline} and \textsc{EgressPipeline} are considered,
as almost all other components are minimally specified, and the packet
replication engine is treated as a function that can be computed. The
execution semantics of the P4-programmable components are instantiated
by the specific P4 sampler program. We specified the required behavior
of the \textsc{IngressPipeline} and proved that the P4 program
satisfies that spec using the Verifiable P4 program logic, did the
same for the \textsc{EgressPipeline}, then used those lemmas to finish
stage 1. Essentially, a typical lemma for a P4-programmable component
takes the following form:
\begin{equation}\label{eqn:triple}
  \textsc{Precondition}(d, s) \rightarrow
  R_c(d, s, d', s') \rightarrow
  \textsc{Postcondition}(d', s')
\end{equation}
Unfolding the layers of definitions within \textsc{IngressPipeline}
and \textsc{EgressPipeline} reveals six execution semantics predicates
with the form $R_c(d, s, d', s')$, serving as assumptions. Each
predicate corresponds to a specific P4-programmable component. By
applying the component lemmas, in the form of \eqref{eqn:triple}, to
these predicates, we can derive the conclusion
$\textsc{Postcondition}(d', s')$ while identifying
$\textsc{Precondition}(d, s)$ as a proof obligation. The
connections between $d'$ and $d$ in $R_c$ enable the derivation of
$\textsc{Precondition}(d, s)$ based on the $\textsc{Postcondition}(d',
s')$ from upstream components. This methodological chaining is crucial
for linking all components. Once all proof obligations are fulfilled,
the conclusions drawn from each component's lemma facilitate the
verification of the desired property. Throughout the proof process,
it's common to encounter mismatches between the postconditions and
preconditions that were assumed and proved for each component
independently.  That is, the chaining proofs together may expose flaws
in these specifications, which are then corrected in the
verification-engineering process.  Typically, one must introduce and
prove additional properties in the specification. Incorporating these
properties allows us to effectively link all components, culminating
in the establishment of \textsc{Normal\-Packet\-Relation} and
\textsc{Special\-Packet\-Relation}.

The proof in Stage 2 is relatively straightforward compared to Stage
1. Using the axioms from Section~\ref{sec:configurable}, we can
eliminate most configurable engines. By performing induction on the
length of the input queue, we can prove the conclusion of
Theorem~\ref{thm:sampler}.

\subsection{Stateful firewall}
Wang \emph{et al.} \cite{verifiablep4} used the Verifiable P4 program logic
and tool to prove correctness of a stateful firewall.  Correctness, as stated and proved, 
relies on the assumption that the incoming packet flow
(from all sources combined)
has no
gaps lasting longer than 10 milliseconds.  One can configure the Tofino
packet generator to insert a packet into the (combined) flow every 10 ms,
which provably discharges this assumption---if one has a specification
of the packet generator and its connection to the P4 program,
which we now have.  Therefore we have proved that our given configuration
of the packet generator satisfies that precondition for correctness
of the stateful firewall.

\section{Related Work}\label{sec:relatework}

\subsubsection*{Verified Parsers.} Over the years researchers have developed quite a
few instances of verified parsers or parser
interpreters/generators. Blaudeau and Shankar~\cite{packrat} present a
formalized metatheory of Parsing Expression Grammars (PEGs) and
develop a verified reference parser interpreter, a packrat parser and
a semantic interpreter, all implemented and proved correct within the
PVS specification and verification system. Delaware \emph{et
al.}~\cite{narcissus} present \textsc{Narcissus}, a framework for
deriving provably correct-by-construction decoders and encoders from
binary format specifications, using the Coq proof assistant to ensure
their correctness and to automate their generation through verified
combinators. Our format notations are inspired by their binary format
specifications. Doenges \emph{et al.}~\cite{leapfrog} introduce a
Coq-based framework for verifying the equivalence of network protocol
parsers. This framework uses an automata model for P4 parsers and an
algorithm for symbolically computing a compact representation of a
bisimulation, which are then verified through a certified compilation
chain, making the parser-equivalence proofs fully automatic. Van Geest
and Swierstra~\cite{packetdesc} introduce a method for describing
binary data formats using a collection of data types embedded in a
dependently typed programming language, which enables automatic
generation and verification of parsers and pretty printers to ensure
correctness by construction, demonstrated through a case study of the
IPv4 network protocol. Ramananandro \emph{et al.}~\cite{everparse}
introduce EverParse, a framework for generating secure, memory-safe,
and non-malleable parsers and serializers from binary message format
descriptions, verified in F*. They demonstrate the framework's
effectiveness through case studies on TLS, Bitcoin, and PKCS \#1,
showing improved performance and security. Ye and
Delaware~\cite{protocolbuf} develop a formally verified compiler for a
subset of the Protocol Buffer serialization format, proving the
correctness of generated serializers and deserializers using the Coq
proof assistant. As we can see, all these works involve verified
parser generators and compilers based on different specifications
(PEGs, binary format specifications, serialization formats).
In future work, we may retarget one of these parser-synthesis
techniques to P4, producing formally proved correct-by-construction
parsers.

\subsubsection*{P4 Architecture Formalizations.}
Wang \emph{et al.}~\cite{verifiablep4} presented a formal
big-step semantics for the P4 programming language;
Alshnakat \emph{et al.}~\cite{hol4p4_2024} presented a formal
small-step semantics, capturing all
key features including concurrency and externs, and providing a
formalized type system and verification framework within the HOL4
theorem prover. 
Whether a small-step semantics is needed, or whether
a big-step semantics suffices, may depend on architectural assumptions about
interpacket and interpipeline visibility of computation steps.

Alshnakat \emph{et al.} provide an architectural level
semantics instantiated in HOL4 for the VSS, V1Model, and eBPF
architectures (which have much simpler traffic managers than
Tofino). They did not demonstrate how to use these models in program
verification---only in simulation on concrete inputs.

\subsubsection*{P4 Verifiers.}
Wang \emph{et al.} \cite[\S 7]{verifiablep4} and Alshnakat \emph{et al.} \cite[\S 5]{hol4p4_2024}
both discuss prior work in mechanized verifiers for P4 programs.
Our contribution here is not a new P4 verifier, but a way to place
such a verifier into a broader context to prove more
comprehensive theorems about packet-stream processors.

\section{Conclusion and Lessons Learned}
\label{sec:conclusion}

Whole-architecture, whole-program, whole-stream-property correctness
proofs for switch programming and configuration can be practical
if it specifications and proofs are sufficiently modular.
We achieved the goal of separating the P4 program-correctness proof from the 
specifications and proofs of configurable engines.
Furthermore, the modularity of configurable-engine components
successfully transfers into specifications and proofs.
But there are some
difficulties:
\begin{itemize}
\item A goal of the Open Networking Foundation
  was to standardize the P4 language \cite{p416}.
  Therefore, (at least in principle) tools such as
  Verifiable P4 can be reusable on other
  P4-programmable architectures.  But no effort was ever made to standardize
  the other ``configurable engines''.  Therefore the \emph{approach} taken
  in this work may be portable and retargetable, but the particular
  specifications are not, and nor will be the programs and proofs of
  application such as the packet sampler.
\item P4 programming permits partially (un)initialized records,
  and the Verifiable P4 logic has a general and elegant treatment of them
  \cite[\S 3.2]{verifiablep4}.  But there is little use for uninitialized
  bits in parsers, deparsers, and engines, so at the boundary between
  P4 and the engines we must constantly prove initialization properties,
  and this reasoning is quite tedious.  Furthermore, statelessness properties
  such as \textsc{LangSec} become more difficult to formalize when some bits
  can ``float free''--we could not prove this without modifying the program
  to initialize some (supposedly irrelevant) bits.  In light of that, it is not at 
  all clear that this feature of the P4 language is worthwhile.
\item In proving a P4 program that inspects and modifies only a few
  sub-subfields of some headers, our proofs regarding the preservation
  of the unchanged fields were less than elegant; we need better
  abstraction techniques.
\item Our axiomatizations of the configurable engines are, we believe,
  conservative approximations: we have not axiomatized all their features.
  And we have not specified the details of their internal queue sizes,
  which means it's not possible to prove certain guarantees about
  when packets will not be dropped.  Internet architecture
  is designed to tolerate a few dropped or locally reordered packets
  as a tradeoff for extremely high throughput, so network switches
  do not attempt to guarantee never-drop or never-reorder properties,
  nor is it useful to do so.
\end{itemize}

\begin{acks}
This material is based upon work supported by DARPA Contract HR001120C0160.
%% \subsubsection*{Disclosure of Interests}
%% The authors have no competing interests to declare that are
%% relevant to the content of this article. 
\end{acks}
\bibliographystyle{plain}
\bibliography{p4}

\end{document}

%% file: paths.tex
\begin{tikzpicture}[
  block/.style={shape=rectangle, draw=black, fill=blue!20, align=center, rounded corners,
    minimum width=2cm, minimum height=1.3cm},
  fixed/.style={shape=rectangle, draw=black, fill=gray!20, align=center, rounded corners,
    minimum width=2cm, minimum height=1.8cm},
  to/.style={->, >={Stealth[length=8pt]}, thick},
  extern/.style={shape=rectangle, draw=black, fill=orange!30, align=center, rounded corners,
    minimum width=2cm, minimum height=1.3cm},
  decoration={markings, mark=at position 0.5 with {\arrow{Stealth[length=8pt]}}}]
  \node[block] (ingress parser) {Ingress\\Parser};
  \node[block, right=4cm of ingress parser] (ingress) {Ingress\\Control};
  \node[below] at (ingress.south) {(Match-Action Pipeline)};
  \node[block, right=4cm of ingress] (ingress deparser) {Ingress\\Deparser};
  \node[fixed, below=2cm of ingress deparser] (mirror lookup) {Mirror\\Session\\Lookup};
  \node[fixed] at (ingress parser |- mirror lookup) (packet scheduler) {Packet\\Scheduler};
  \node[fixed] at ($(packet scheduler)!.4!(mirror lookup)$) (packet replication) {Packet\\Replication\\Engine};
  \node[fixed] at ($(packet scheduler)!.6!(mirror lookup)$) (write admission) {Write\\Admission\\Control};
  \node[fixed] at ($(packet replication)!.5!(packet scheduler)$) (queue admission) {Queue\\Admission\\Control};
  \node[fixed] at ($(write admission)!.5!(mirror lookup)$) (mirror buffer) {Mirror\\Buffer\\Merge};
  \draw[to] (ingress parser) -- node [above, align=center] {Intrinsic Meta Data\\Custom Meta Data\\Header} (ingress);
  \draw[to] (ingress) -- node [above, align=center] {Intrinsic Meta Data\\Custom Meta Data\\Header} (ingress deparser);
  \draw[to] (mirror lookup) -- (mirror buffer);
  \draw[to] (mirror buffer) -- (write admission);
  \draw[to] (write admission) -- (packet replication);
  \draw[to] (packet replication) -- (queue admission);
  \draw[to] (queue admission) -- (packet scheduler);
  \node[block, below=2cm of packet scheduler] (egress parser) {Egress\\Parser};
  \node[block] at (ingress |- egress parser) (egress) {Egress\\Control};
  \node[below] at (egress.south) {(Match-Action Pipeline)};
  \node[block] at (mirror lookup |- egress) (egress deparser) {Egress\\Deparser};
  \node[label=left:CPU] at ($(ingress deparser)+(0.5cm, 2.7cm)$) (CPU) {\includegraphics[height=1cm]{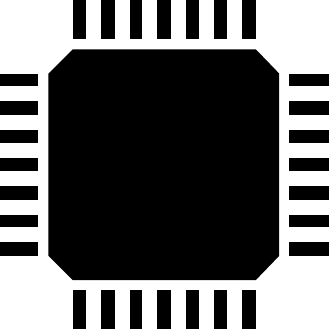}};
  \node[extern, below=1.5cm of egress deparser] (output ports) {Output\\Ports};
  \node[extern, left=1cm of ingress parser] (input ports) {Input\\Ports};
  \node[fixed, below=2cm of input ports] (packet gen) {Packet\\Generator};
  \draw[to] (input ports) -- (ingress parser);
  \draw[to] (packet gen) -- node[right] {Packet} (input ports);
  \draw[to] ($(ingress deparser.north)+(0.5cm, 0)$) -- (CPU);
  \draw[to] (egress parser) -- node [above, align=center] {Intrinsic Meta Data\\Custom Meta Data\\Header} (egress);
  \draw[to] (egress) -- node [above, align=center] {Intrinsic Meta Data\\Custom Meta Data\\Header} (egress deparser);
  \draw[to] ($(ingress deparser.south)+(0.5cm, 0)$) -- node [left, near end] {Mirrored Packet} ($(mirror lookup.north)+(0.5cm, 0)$);
  \draw[to] (packet scheduler) -- node [right, align=left] {Packets} (egress parser);
  \draw[to] (egress deparser) -- node [left] {Mirrored Packet} (mirror lookup);
  \coordinate (IDU) at ($(ingress deparser)+(0, 2cm)$);
  \coordinate (IPU) at ($(ingress parser)+(0, 2cm)$);
  \draw[thick] (ingress deparser) -- (IDU);  
  \draw[thick, postaction={decorate}] (IDU) -- node[above] {Resubmitted Packets} (IPU);
  \draw[to] (IPU) -- (ingress parser);
  \node at ($(ingress parser)!.5!(packet scheduler)$) (bin){\includegraphics[height=1cm]{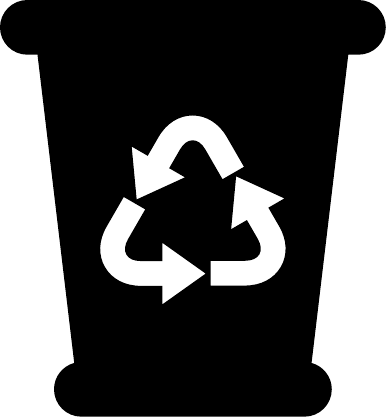}};
  \draw[to] (ingress parser) -- (bin);
  \coordinate (binlo) at ($(bin.east)-(0, 0.2cm)$);
  \coordinate (binhi) at ($(bin.east)+(0, 0.2cm)$);
  \draw[thick] (queue admission) -- (queue admission |- binlo);
  \draw[to] (queue admission |- binlo) -- (binlo);
  \coordinate (IDD) at (ingress deparser |- binhi);
  \draw[thick] (ingress deparser) -- (IDD);
  \draw[thick] (IDD) -- node[above] {Normal Packet} (mirror buffer |- IDD);
  \draw[to] (mirror buffer |- IDD) -- (mirror buffer);
  \draw[thick] (write admission) -- (write admission |- binhi);
  \draw[to] (write admission |- binhi) -- node[above, near end] {Dropped Packets} (binhi);
  \draw[to] (egress deparser) -- node[left] {Normal Packets} (output ports);
  \coordinate (PSL) at ($(packet scheduler)+(-1.5cm,0)$);
  \coordinate (OPU) at ($(output ports.west)+(0,0.3cm)$);
  \draw[thick] (packet scheduler) -- (PSL);
  \draw[thick, postaction={decorate}] (PSL) -- (PSL |- OPU);
  \draw[to] (PSL |- OPU) -- node[above] {Bypass Packet} (OPU);
  \coordinate (OPB) at ($(output ports.west)+(0,-0.3cm)$);
  \draw[thick, postaction={decorate}] (OPB) -- node[above] {Recirculation} (packet gen |- OPB);
  \draw[to] (packet gen |- OPB) -- (packet gen);
  \draw[thick, dashed]($(packet scheduler.south west)-(0.2cm, 0.2cm)$) rectangle ($(mirror lookup.north east)+(0.2cm, 0.2cm)$);
  \node at ($(egress)+(0,2.1cm)$) {Traffic Manager};
\end{tikzpicture}

%% file: pre.tex
\begin{tikzpicture}
  [block/.style={shape=rectangle, draw=black, align=center, minimum
      width=2cm, minimum height=1.3cm},
    ->/.style={black, -Stealth},
    que/.style={rectangle split, rectangle split parts=#1, rectangle split horizontal,
      minimum height=0.5cm, draw}]
  \node[block, rounded corners] (unicast) {Unicast\\Engine};
  \node[block, rounded corners, below=0.5cm of unicast] (multicast) {Multicast\\Engine};
  \node[block, rounded corners, below=0.1cm of multicast] (replicator) {Packet\\Replicator};
  \node[block, fill=green!20, anchor=east] (packet) at
  ($(unicast.north west)!.5!(multicast.south west)-(1.0cm,0)$) {Packet\\with Metadata};
  \node[block, left=1cm of replicator] (config) {Multicast\\Group\\Configuration};
  \draw[thick] (replicator) -- (multicast);
  \draw[->, out=0, in=180] (packet) to (unicast);
  \draw[->, out=0, in=180] (packet) to (multicast);
  \draw[->, out=0, in=180] (config) to (multicast);
  \node[que=10, anchor=west, label=below:Packet Queue,
    rectangle split part fill={white, white, white, green!20, green!20, green!20,
      green!20, green!20, green!20, white}] (queue) at
  ($(unicast.north east)!.5!(multicast.south east)+(1.0cm,0)$) {};
  \draw[->, out=0, in=180] (unicast) to (queue);
  \draw[->, out=0, in=180] (multicast) to (queue);
  \draw[->] (queue.east) -- +(1cm, 0);
\end{tikzpicture}

%% file: groupconfig.tex
\begin{tikzpicture}
  [col/.style={rectangle split, rectangle split parts=#1, draw},
    ->/.style={black, -Stealth}]
  \node[col=4, label=above:Level 1 Nodes] (level1 nodes1)
       {Node 1\nodepart{two}Node 2\nodepart{three}Node 3\nodepart{four}\ldots};
  \node[col=4, below=0.5cm of level1 nodes1] (level1 nodes2)
            {Node 1\nodepart{two}Node 2\nodepart{three}Node 3\nodepart{four}\ldots};
  \node[col=5, anchor=east] (group ids) at
  ($(level1 nodes1.north west)!.5!(level1 nodes2.south west)-(1.0cm,0)$)
       {$\mathsf{ID_1}$\nodepart{two}$\mathsf{ID_2}$\nodepart{three}
         $\mathsf{ID_3}$\nodepart{four}\ldots\nodepart{five}$\mathsf{ID_n}$};
  \draw[->, out=0, in=180] (group ids.text east) to (level1 nodes1.north west);
  \draw[->, out=0, in=180] (group ids.two east) to (level1 nodes2.north west);
  \node[col=6, anchor=west] (node fields) at
  ($(level1 nodes1.north east)!.5!(level1 nodes2.south east)+(1.0cm,0)$)
       {\verb|dev_port_list|\nodepart{two}\verb|lag_list|
         \nodepart{three}\verb|L1_XID_VALID|\nodepart{four}\verb|L1_XID|
         \nodepart{five}\verb|RID|\nodepart{six}\ldots};
  \draw (level1 nodes1.two split east) -- (node fields.north west);
  \draw (level1 nodes1.three split east) -- (node fields.south west);
  \node[col=10, anchor=west, right=1cm of node fields, label=above: Level 2 Nodes]
  (port list)
       {Port 1\nodepart{two}Port 2\nodepart{three}Port 3\nodepart{four}\ldots
         \nodepart{five}Port $m$\nodepart{six}LAG 1\nodepart{seven}LAG 2
         \nodepart{eight}LAG 3\nodepart{nine}LAG 4\nodepart{ten}\ldots};
  \draw (node fields.north east) -- (port list.north west);
  \draw (node fields.two split east) -- (port list.south west);
\end{tikzpicture}